\documentclass[a4paper,fleqn,amssymb,usenatbib,useAMS,usedcolumns]{mnras}
\usepackage{times} 
\usepackage{latexsym,amssymb,amsmath,amsfonts}
\usepackage{comment}
\usepackage{rotating} 
\usepackage{float}
\usepackage{graphicx}
\usepackage{longtable}
\usepackage{url}
\usepackage{dcolumn}
\usepackage{textcomp}
\usepackage{multicol}
\usepackage{bm}
\usepackage{pdflscape}
\usepackage[T1]{fontenc}
\usepackage{ae,aecompl}
\usepackage[caption=false]{subfig}
\usepackage[dvipsnames]{xcolor}
\usepackage{lscape}
\usepackage{afterpage}
\usepackage{pdflscape}
\usepackage{hyperref}
\usepackage{listings}
\usepackage{adjustbox}
\usepackage[normalem]{ulem}

\newcommand{\HI}{\mbox{H\,{\textsc i}}}
\newcommand{\zabs}{$z_{\rm abs}$}

\newcommand{\cmsq}{cm$^{-2}$}
\newcommand{\degree}{\ensuremath{^\circ}}
\newcommand{\Msun}{$M_{\odot}$}

\newcommand{\hi}{\ion{H}{\sc i}}
\newcommand{\mhi}{$M_\ion{H}{i}$}

\newcommand{\kms}{$\rm{km\, s^{-1}}$}

\newcommand{\nai}{\mbox{Na\,{\sc i}}}
\newcommand{\caii}{\mbox{Ca\,{\sc ii}}}
\newcommand{\carcsec}{$^{\prime\prime}$}
\newcommand{\kmps}{$\mathrm{km\,s^{-1}}$}

\usepackage{threeparttable}
\usepackage{longtable}

\title[Klemola\,31 group in emission and absorption]{Mapping \hi\ 21-cm in the Klemola 31 group at $z = 0.029$: emission and absorption towards PKS\,2020-370}

\author[E. K. Maina et al.]{
E. K. Maina$^{1}$\thanks{E-mail: kamau757@gmail.com},
Abhisek Mohapatra,$^{2}$
G. I. G. J\'ozsa,$^{3,1}$
N. Gupta,$^{2}$
F. Combes,$^{4}$
P. Deka,$^{2}$
\newauthor
J. D. Wagenveld,$^{3}$
R. Srianand,$^{2}$
S. A. Balashev,$^{5,6}$
Hsiao-Wen Chen,$^7$
J.-K. Krogager,$^8$
E. Momjian,$^{9}$
\newauthor
P. Noterdaeme,$^{10,11}$ and
P. Petitjean$^{10}$
\\
$^{1}$Department of Physics and Electronics, Rhodes University, P.O. Box 94, Makhanda, 6140, South Africa \\ 
$^{2}$Inter-University Centre for Astronomy and Astrophysics, Post Bag 4, Ganeshkhind, Pune 411 007, India\\
$^{3}$Max-Planck-Institut f\"ur Radioastronomie, Auf dem H\"ugel 69, D-53121 Bonn, Germany\\
$^4$  Observatoire de Paris, LERMA, College de France, CNRS, PSL University, Sorbonne University, Paris, France\\
$^5$ Ioffe Institute, Politekhnicheskaya 26, 194021 Saint Petersburg, Russia\\
$^6$ HSE University, Saint Petersburg, Russia \\
$^7$ Department of Astronomy \& Astrophysics, The University of Chicago, 5640 South Ellis Avenue, Chicago, IL 60637, USA \\
$^8$ Université Lyon1, ENS de Lyon, CNRS, Centre de Recherche Astrophysique de Lyon UMR5574, F-69230 Saint-Genis-Laval, France\\
$^{9}$ National Radio Astronomy Observatory, P.O. Box O, Socorro, NM 87801, USA\\
$^{10}$ Institut d’Astrophysique de Paris, Sorbonne Université and CNRS, 98bis boulevard Arago, F-75014 Paris, France\\
$^{11}$ Franco-Chilean Laboratory for Astronomy, IRL 3386, CNRS and U. de Chile, Casilla 36-D, Santiago, Chile\\
}

\date{Accepted XXX. Received YYY; in original form ZZZ}

\pubyear{2022}

\begin{document}
\label{firstpage}
\pagerange{\pageref{firstpage}--\pageref{lastpage}}
\maketitle

\begin{abstract}
We present MeerKAT Absorption Line Survey (MALS) observations of the \hi\ gas in the Klemola\,31 galaxy group ($z=0.029$), located along the line of sight to the radio-loud quasar PKS\,2020-370 ($z=1.048$). Four galaxies of the group are detected in \hi\ emission, and \hi\ absorption is also detected in front of PKS\,2020-370 in Klemola\,31A. The emission and absorption are somewhat compensating on the line of sight of the quasar, and the derived column density of the absorption appears under-estimated, with respect to the neighbouring emission. A symmetric tilted-ring model of Klemola\,31A, assuming the absorbing gas in regular rotation in the plane, yields a rather high spin temperature of 530 K. An alternative interpretation is that the absorbing gas is extra-planar, which will also account for its non-circular motion. 
The \nai/\caii\ ratio also suggests that the absorbing gas is unrelated to cold \hi\ disk. Two of the galaxies in the Klemola group are interacting with a small companion, and reveal typical tidal tails, and velocity perturbations. Only one of the galaxies, ESO\,400-13, reveals a strong \hi\ deficiency, and a characteristic ram-pressure stripping, with a total asymmetry in the distribution of its gas. Since a small galaxy group as Klemola\,31 is not expected to host a dense intra-group gas, this galaxy must be crossing the group at a very high velocity, mostly in the sky plane.

\end{abstract}




\section{Introduction}
\label{sec:intro}

Atomic hydrogen or \hi\ is a transitional phase of the most abundant atom in the universe. On its path from ionized to molecular hydrogen (H$_2$), the basic fuel for star formation, it must transit through a cool phase forming the cold neutral medium (CNM; $T\sim$100\,K). Thus, the CNM fraction of gas in a galaxy is expected to be related to the efficiency with which gas is converted into stars. On the other hand, the warmer circumgalactic gas, which eventually might fuel star formation, also has a neutral component, which can be traced using \hi\ observations.  Vice versa, gas in galaxies being ionized from feedback processes must also go through this phase and might be pushed into the outskirts of galaxies. \hi\ is therefore also a tracer of gas depletion. Studies of \hi\ in the interstellar and intergalactic medium are hence of utmost importance to understand the processes that govern galaxy formation and evolution.

In the nearby Universe, the \hi\ 21-cm emission line observations can be used to map the distribution and kinematics of bulk neutral gas, i.e., regardless of the kinetic temperature of gas, associated with galaxies.  \hi\ 21-cm emission line observations are primarily limited by resolution effects and the column density detection limit increases with the distance.
The availability of resolved \hi\ 21-cm line imaging of large number of galaxies from the ongoing large survey projects with the Square Kilometer Array (SKA) precursor telescopes i.e., the Australian Square Kilometre Array Pathfinder (ASKAP) \citep[][]{Hotan21} and MeerKAT \citep[][]{Jonas16} offers an unprecedented opportunity to address these issues \citep[e.g.,][]{Jarvis17,Serra16, Koribalski20,maddox_mightee-hi_2021}.
This is, however, with the exception of very deep observations providing a low-number statistics of resolved galaxies, mostly still restricted to a redshift range of $z\,<\,0.1$.  
At higher redshift, the neutral gas can in turn be detected through \hi\ 21-cm absorption whenever it lies between the observer and a strong background radio source.
This technique, however, generally provides information along one sight line, albeit at a resolution defined by the angular size of the background radio continuum. 
Another important characteristic of this technique is that the strength of the \hi\ 21-cm absorption is inversely proportional to the spin temperature ($T_{\rm s}$) of the gas and, hence, more efficient in tracing the colder atomic phases \citep[e.g.,][]{Kulkarni88}. \hi\ 21-cm absorption and emission line observations therefore nicely complement each other.

To interpret \hi\ absorption at higher redshift, it is important to conduct studies at lower redshift, where both emission and absorption signals are detectable.
It is now well recognized that mergers and interactions with other galaxies, and ram pressure striping due to hot gas in group and cluster environments, play an important role in defining the overall structure i.e., formation of arms and bars, and the gas content of galaxies \citep[e.g.,][]{Dressler1980351,Gerin90, Elmegreen91, Irwin94,Yun94,Cappellari2011,reynolds_wallaby_2021,wang_wallaby_2021,Castignani22}. 
These processes will feed the circumgalactic medium, and it is not yet quite clear which of these media is actually traced by \hi\ absorption lines in general.  The low-redshift \hi\ 21-cm emission and absorption line observations can help in recognizing the processes dominating in different galaxy environments.

The large survey project {\it MeerKAT Absorption Line Survey} (MALS) is carrying out a sensitive dust-unbiased search of \hi\ 21-cm and OH 18-cm absorption lines with a primary goal to better understand the evolution of cold atomic and molecular gas at $0<z<2$ \cite[][]{Gupta2016}. 
Owing to the large field-of-view ($\sim1^{\circ}$ at 1.4\,GHz)  and excellent sensitivity, MALS can also detect \hi\ 21-cm line emission from nearby galaxies.  The background radio loud AGN within the MALS pointing can then be used to simultaneously search for \hi\ 21-cm absorption in galaxies and the associated environment. An excellent demonstration of this is presented by \citet[][]{Boettcher22} reporting the discovery of neutral gas detected in both damped Ly$\alpha$ absorption (DLA) and \hi\ 21-cm emission outside of the stellar body of a dwarf galaxy, at a distance of 33\,kpc from another dwarf galaxy.  The interaction between these two dwarf galaxies is revealed by a bridge detected in \hi\ emission, and the non-detection of \hi\ 21-cm absorption shows that the gas producing DLA is warm ($T_{\rm s}>$1880\,K) or clumpy. 
%
Overall, MALS offers obvious advantages of combining \hi\ 21-cm emission and absorption line measurements in revealing the physical state and parsec-scale structure of atomic gas in nearby galaxies, which till now have been scarcely applied beyond the Milky Way \citep[e.g.,][]{Carilli92, Borthakur14, Dutta16, Reeves16, Gupta18}.  

In this work, we present a detailed \hi\ 21-cm absorption and emission line study of the Klemola\,31 galaxy group ($z_{grp}=0.029$) in the foreground of a distant quasar, PKS\,2020-370 (RA: 20:23:46.200, Dec -36:55:20.50 \citep[J2000;][]{Healey08} at $z_q$ = 1.048 \citep[][]{Peterson76}. 
The group consists of six galaxies and the quasar sightline passes through one of them i.e., Klemola\,31A. Previously, detections of  \ion{Ca}{II}, K and H lines \citep[][]{Boksenber1980} and a faint \hi\ 21-cm absorption \citep[][]{Boisse88, Carilli87} have been reported in the literature. Through MeerKAT, we detect \hi\ 21-cm emission from the galaxy group and absorption associated with Klemola\,31A. We also complement the latter with the upgrade Giant Metrewave Radio Telescope (uGMRT).

%
\begin{table*}
\label{tab:group_gal} 
    \centering
\caption{List of galaxy members of Klemola\,31 group.}
\small{
\begin{tabular}{cccccccccc}
\hline
\hline
Name  & RA & DEC & $z_g^{+}$ & ${\rm{B\,mag}}^\dag$ & $\rm D_{25}^\dag$& $i_{opt}^\dag$ & Morph.$^\dag$ & Angular & Impact \\
& & & & & & & & separation & parameter\\
& & &  &(mag) & (kpc) &  & & (arcmin) & (kpc) \\
(1) & (2) & (3) & (4) & (5) & (6) & (7) & (8) &(9) &  (10) \\
\hline \\
ESO 400-11 & 20:23:36.81 & -36:52:55.48 & 0.0284  & 15.79 & $42.7\pm 4.0$& $79^{\circ}$ &Sc & 2.75 & 96\\
ESO 400-13  & 20:24:01.94 & -36:56:16.99  & 0.0292 & 14.98 & $41.6\pm4.2$ & $64^{\circ}$ &S0 & 3.55 &124 \\
Klemola\,31A  & 20:23:45.21 & -36:55:05.87  & 0.0288 & $15.91$ & $31.5\pm3.9$ & $68^{\circ}$ &Sc & - & - \\
LEDA\,631562  & 20:24:06.06 & -36:52:24.00  & 0.0292 & 16.52 & $21.5\pm4.3$ & $46^{\circ}$ &Sbc$^*$ & 5.06 & 176 \\
Klemola\,31B & 20:23:50.06 & -36:55:18.54 & 0.0286 & 16.80 & $19.2\pm6.7$& $55^{\circ}$ &E$^*$ & 0.99 & 35\\
LEDA\,2807038  & 20:23:34.76 & -36:53:15.31  & 0.0295 & 16.80&$15.2\pm6.3$& $39^{\circ}$ &E$^*$ & 2.79 & 97 \\

\hline
\end{tabular}
}
\begin{tablenotes}
\small
    \item Columns\,$1-3$: Galaxy name, position i.e., right ascension and declination (J2000) from \textsc{SIMBAD}\footnote{\href{http://simbad.u-strasbg.fr/simbad/}{http://simbad.u-strasbg.fr/simbad/}}. Column\,4: redshift of the galaxy obtained from the optical spectra of 6dF Galaxy Survey (6dFGS); for Klemola31A see \citet[][]{Boksenber1980} and \citep[][]{Carilli87}. Column\,5: Total B-band magnitude.
    Column\,6: the extinction-corrected optical diameter measured at the 25th mag arcsec$^{-2}$ isophote from HyperLeda. Column\,7: inclination angle calculated from the axis ratio of the isophote at 25 mag arcsec$^{-2}$ in the  B-band. 
    Column\,8: Morphological type of the galaxy. Column\,9: Angular separation from Klemola\,31A. Column\,10: Projected distance from Klemola\,31A using average redshift of the group, $z_{grp} = 0.0290$. \\
    $^\dag$from \textsc{Hyperleda}; $^*$ from visual classification based on the optical image.
\end{tablenotes}
\end{table*}

This paper is organized as follows. In Section~\ref{sec:pksklem}, we introduce the properties of Klemola\,31 group. In Section~\ref{sec:obs}, we present MeerKAT and uGMRT observations, and data analysis. In Section~\ref{sec:res}, we present  details  of the \hi\ 21-cm emission line properties of the group followed by the \hi\ 21-cm absorption line properties towards PKS\,2020-370 and the implications. A summary of results is presented in Section~\ref{sec:summ}. Throughout the paper, we use $\Lambda CDM$ cosmology with $H_0 =70$ {\kms}\,Mpc$^{-1}$, $\Omega_\Lambda$ = 0.7, and $\Omega_m$ = 0.3. At the distance of the Klemola\,31 galaxy group ($z_{grp}=0.029$),  $1^{\prime\prime}$ = 0.6\,kpc.

\section{Klemola\,31 group}
\label{sec:pksklem}

Klemola\,31 is a small group of galaxies at $z_{grp}=0.029$ located in the Sagittarius constellation in the southern hemisphere \citep[][]{Klemola69}. The group consists of six galaxies with different masses, morphological types and gas distributions. The basic properties of the member galaxies are summarized in Table~\ref{tab:group_gal}.
The redshifts of five of the six member galaxies are based on the publicly available optical spectra from the 6dF Galaxy Survey (6dFGS)\footnote{\href{http://www.6dfgs.net/}{http://www.6dfgs.net/}}. In short, the redshifts of ESO\,400-11 and LEDA\,631562 are based on the prominent $\rm H\alpha$ ($\lambda6565$ \AA) emission line.  For the early type galaxies, ESO\,400-13, Klemola\,31B and LEDA\,2807038, the Mg ($\lambda5177$ \AA) and Na ($\lambda5896$ \AA) absorption lines have been used. For Klemola\,31A ($z_g = 0.0288$), which has no public spectra available, the redshift is taken from \citet[][]{Boksenber1980} and \citet[][]{Carilli87}.
With respect to Klemola\,31A, the other members of the group are within an angular separation of $\sim5\arcmin$, which corresponds to a projected distance of $\sim$176\,kpc at the mean redshift of the group, $z_{grp}$ = 0.029 (see last column of Table~\ref{tab:group_gal}). The background quasar PKS\,2020-370 is at an angular separation of $18\farcs7$, i.e., 10\,kpc at $z_{grp}$, from Klemola\,31A.

\citet[][]{Boisse88} detected \HI\ emission from the group using the Nancay radio telescope.  The telescope beam covers four members of the group: Klemola\,31A, Klemola\,31B,  ESO\,400-11 and LEDA\,2807038.  Using the morphology and inclinations of these galaxies,  \citet[][]{Boisse88} argued that the detected \hi\ emission signal (integrated flux, $S=0.78$ Jy\,\kms) corresponds to Klemola\,31A (\mhi = $6 \times 10^9\, h^{-2}$).  
Later, \citet{Carilli92} using the Very Large Array (VLA of the NRAO\footnote{The National Radio Astronomy Observatory is a facility of the National Science Foundation operated under cooperative agreement by Associated Universities, Inc.}) in BnA and CnB configurations constrained the \HI\ content of Klemola\,31A and Klemola\,31B to be \mhi = $1.16 \times 10^9\, h^{-2}$ \Msun\ and \mhi\ $<10^9\, h^{-2}$ \Msun, respectively.  For the \hi\ mass upper limit they assumed a velocity width of 200 \kmps.

Through the detection of $\ion{Ca}{II}$ K and H absorption lines at \zabs $=0.0287$ in the spectra of PKS\,2020-370 \citep[][]{Boksenber1980}, it has been known for a long time that gas extends well outside the optical disks / halos of Klemola\,31A and Klemola\,31B.  The latter is at an angular separation of 45$\arcsec$ from the quasar sight line. The above mentioned \hi\ 21-cm emission line observations subsequently showed the gas to be associated with the extended \hi\ disk of Klemola\,31A. \citet[][]{Junkkarinen94} detected Na~{\sc i} absorption lines and showed that the ratio $N$(Ca~{\sc ii})/$N$(Na~{\sc i}) is rather consistent with the `halo-like' gas \citep[see also][]{Cherinka11}.  Both \citet[][]{Boisse88} and \citet[][]{Carilli92} have reported the detections of \hi\ 21-cm absorption from  Klemola\,31A.  However, the best available constraint on integrated \hi\ 21-cm optical depth ($\int\tau$dv = $0.22\pm0.05$\,\kms) is based on VLA A-configuration observations (spectral resolution $\sim$10\,\kms per channel) that resolve out the extended \hi\ emission signal \citep[][]{Carilli87}.  This corresponds to a \hi\ column density, $N$(\hi) = $4.0 \times 10^{19} (T_{s} / 100\, \rm{K} )$ cm$^{-2}$.

The availability of a variety of \hi\ 21-cm emission and absorption line information motivated the use of Klemola group as target for MALS science verification observations..  At the same time, the \HI\ 21-cm emission line properties of the other four members of the galaxy group i.e., ESO\,400-11, LEDA 2807038, LEDA 631562 and ESO 400-13 have not been discussed at all in the literature, and the absorption signal towards PKS\,2020-370 although unambiguously detected, albeit as a single pixel, is not well characterized due to the low spectral resolution. Thus, it is also an intriguing target to follow-up using the large field-of-view, excellent sensitivity and high spectral resolution of MeerKAT observations presented here.

\section{Observations and analysis}
\label{sec:obs}

\subsection{MeerKAT}
\label{sec:obsmrkt}
\noindent The pointing centered at PKS\,2020-370 was observed on April 1, 2020 using 60 antennas of the MeerKAT-64 array and 32K mode of the SKA Reconfigurable Application Board (SKARAB) correlator. The total duration of these science verification observing run was 184 min. The total on-source time of 56\,mins on PKS\,2020-370 was split into three scans at different hour angles to improve the uv coverage. We observed 3C286 and PKS\,1939-638 for flux density scale, delay and bandpass calibrations, and the compact radio source J2052-3640 was periodically observed for complex gain calibration. The total bandwidth of 856\,MHz centered at 1283.9869 was splited into 32,768 channels.  In the vicinity of the redshifted \hi\ 21-cm line frequency at $z=0.029$ ($\sim$1380\,MHz) the channel width corresponds to a velocity resolution of 5.7\,\kms.  The correlator dump time was 8 seconds and the data were acquired for all four polarization products, labelled XX, XY, YX and YY.  Here, we are interested only in the Stokes-$I$ properties of the target,  therefore only XX and YY  polarization products were processed through the Automated Radio Telescope Imaging Pipeline \citep[{\tt ARTIP}; see][for details of processing steps]{Gupta21}.

The spectral line processing through {\tt ARTIP} partitions the frequency band into 15 spectral windows (referred as SPW-0 to -14) with an overlap of 256 frequency channels. For the science objectives of this paper we consider only the continuum images and spectral line cubes from SPW-9 which covers 1351 - 1411\,MHz.  The spatial resolution of the continuum image obtained using {\tt ROBUST = 0} weighting of the visibilities in the CASA {\tt tclean} task is $11\farcs6\times7\farcs.0$ (position angle = $-85.8^\circ$).  The rms noise, measured close to quasar, in this image at the reference frequency of 1380.8\,MHz  is $\sim80\mu$Jy\,beam$^{-1}$.  As expected (see Section~\ref{sec:obsgmrt}), the quasar PKS\,2020-370 is unresolved with a total continuum flux density of 404.1$\pm$1.1\,mJy.  The quoted flux density and the uncertainty are obtained using the single Gaussian component fitted to the continuum image. The typical error on flux density at these low frequencies is expected to be $\sim$5\%.

We also generated a continuum subtracted image cube corresponding to SPW-9. For this we imaged only the central $34\arcmin$ centered at PKS\,2020-370, which is adequate to cover the galaxy group of interest.  This {\tt ROBUST = 0} cube has  1024$\times$1024 spatial pixels of size $2$\carcsec\ and 2304 frequency channels (resolution $\sim$5.7\,\kms).  The image cube has a common restoring beam, i.e. spatial resolution of $11\farcs9\times7\farcs6$ (6.9\,kpc\,$\times$ 4.4\,kpc at $z_{grp}$) 
with a position angle of $-82.7^\circ$), and the rms of 0.79\,mJy\,beam$^{-1}$\,channel$^{-1}$. 
The deconvolution of the line signal present in the image cube is done over two steps.  First, we used the Common Astronomy Software Applications (\textsc{CASA}) package
task {\tt tclean} to deconvolve the signal down to five times the single channel rms.  Then, we used the 3D source finding application, SoFiA\,2\footnote{\href{https://github.com/SoFiA-Admin/SoFiA-2}{https://github.com/SoFiA-Admin/SoFiA-2}} \citep{Serra15, Westmeier2021} to identify the voxels with \HI\ emission signal in the cube. This involved fitting a polynomial of order one to remove the residual continuum subtraction errors, and using the smooth + clip source finder with a combination of spatial kernels of 0, 3, 6, and 9 pixels and spectral kernels of 0, 3, 7, and 15 channels.  A threshold of $3.5\sigma_{rms}$ and the reliability filter with reliability threshold of 0.95 were used to reject unreliable detections.  The output three-dimensional SoFiA mask was then used to further deconvolve the signal down to the single channel rms. This cube was then corrected for the primary beam attenuation using the model from {\tt KATBEAM} library\footnote{\href{https://github.com/ska-sa/katbeam}{https://github.com/ska-sa/katbeam}}. The resultant cube was used for generating \hi\ moment maps and further analysis.  For moment-1 and moment-2 maps, we have used only those pixels where the signal was detected at 1.5 times the local rms.

\subsection{uGMRT}
\label{sec:obsgmrt}
The uGMRT observations were carried out on 2018 May 28 and 29, using a bandwidth of 4.17\,MHz split into 512 channels.  The total on-source time was 6.3\,h.  We also observed 3C\,48 for flux density scale and bandpass calibrations. The compact radio source PKS1954-388 was also observed for 7\,min every $\sim$50\,min.  These observations were planned to complement the MeerKAT observation of PKS2020-370 carried out on 2017, November 7 and 9, using only 16 antennas from the array, with a single objective of detecting the absorption for verification purposes.  These shallow MeerKAT observations are not discussed here, and mentioned only to clarify the motivation of uGMRT observations.

The uGMRT data were also reduced using ARTIP. The Stokes-$I$ image made using {\tt ROBUST = 0} weighting of the visibilities has a restoring beam of $3\farcs9\times 1\farcs8$ with a position angle of $29.4^\circ$.  PKS\,2020-370 is compact at this resolution with a deconvolved size of  $0\farcs7 \times 0\farcs1$ (position angle = $45\pm2^\circ$). The peak and total flux densities are $409.7\pm0.4$\,mJy\,beam$^{-1}$ and $419.3\pm0.8$\,mJy, respectively.  The Stokes-$I$ spectrum of the quasar has a resolution and spectral rms noise of 1.88\,\kms\ and  1.7\,mJy\,beam$^{-1}$\,channel$^{-1}$, respectively.

\section{Results and discussion}
\label{sec:res}

\subsection{Klemola\,31 group in H~{\sc i} emission}
\label{sec:group}
\begin{figure*}
\centering
 \includegraphics[scale=.83]{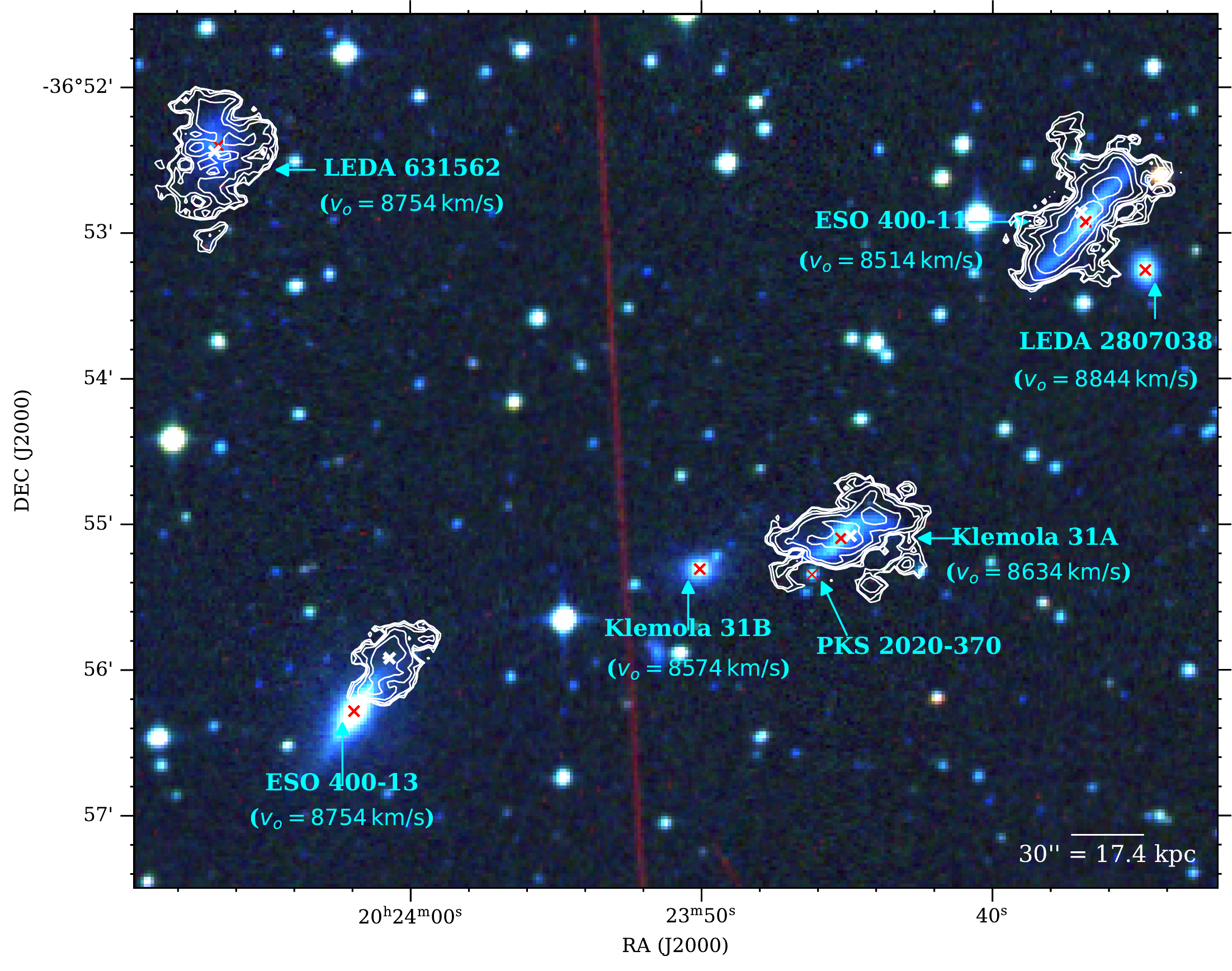}
 \caption{Klemola\,31 group members and PKS\,2020-370 shown in optical
 Digitized Sky Survey (DSS) $B_j$, $R$, and $I$ color composite image with \HI\ column density contours obtained from the moment-0 map. The contours correspond to \HI\ column densities of $ 2^n \times 5 \times 10^{19}$ \,cm$^{-2}$ ($n=0,\,1,\,2,\,3,\, ...)$. The central positions of optical and \HI\ disks are marked using red and white crosses, respectively. The recession velocity ($v_0\, \rm in\,km\,s^{-1}$) corresponding to $z_g$ is denoted below the name of the galaxy. Note that only four of the six group members are detected in \HI\ emission. The radio image has a restoring beam of $11\farcs9\times7\farcs6$ with a position angle of $-82.7^\circ$}.
 \label{fig:Mom0_col_dens}
\end{figure*}

We detect \hi\ 21-cm emission associated with four of the six members of the the Klemola\,31 galaxy group.  In Fig.~\ref{fig:Mom0_col_dens}, we show the DSS  $B_j$, $R$, and $I$ color composite image exhibiting all the group members along with their names obtained from {\tt SIMBAD}.   Also, overlaid on the optical image are \hi\ column density contours based on the MeerKAT data.   
The \hi\ column density is calculated using:
\begin{equation}
\label{eqn:col_dens}
\frac{N_\ion{H}{i}}{\mathrm{cm}^{-2}} = \frac{1.104 \times 10^{24}(1+{\it z})^3}{b_{maj} \times b_{min}} \int \frac{I(v)}{\mathrm{Jy\, beam}^{-1}}\, \frac{dv}{\mathrm{km\,s}^{-1}}.
\end{equation}
\noindent 
Here $b_{maj}$ and $b_{min}$ are major and minor axes of the restoring beam in arcseconds, and $\int \rm{I(v) dv} $ is the total intensity in Jy\,km s$^{-1}\,\mathrm{beam}^{-1}$, with $v$ being measured in the reference frame of the target.  The spectral rms noise in the image cube for an assumed line width of 15\,\kms\ corresponds to the 3$\sigma$ \hi\ column density of $3.1\times10^{20}$\,\cmsq.

The \HI\ emission from the group members resides in the frequency range of $1379.4-1382.2$ MHz. This corresponds to the redshift ($z_\hi$) and recession velocity ($v_o = cz_\hi$) ranges of $0.0276-0.0297$ and $8274.3-8903.8$ \kmps, respectively, estimated from the center of global \hi\ emission profiles of individual galaxies.  The \hi\ masses are calculated using:
\begin{equation}
\label{eqn:Hi_mass}
\frac{M_\ion{H}{i}}{M_\odot} = \frac{2.356 \times 10^5}{1 + z_{\ion{H}{I}}} \times\bigg({\frac{D_L}{\rm{Mpc}}}\bigg)^2\, \int{\frac{S(v)}{\mathrm{Jy}}\,\frac{dv}{\mathrm{km\,s}^{-1}}},
\end{equation}
\noindent
where $D_L$ is the cosmological luminosity distance corresponding to $z_{\ion{H}{I}}$.  All the above mentioned properties, i.e., position, redshift, recession velocity, integrated line flux and \hi\ mass are provided in rows 1 - 6 of  Table~\ref{tab:HI_properties}.  The 3$\sigma$ upper limits on \hi\ masses ($<5\times10^8$\,\Msun) of galaxies undetected in \hi\ emission, i.e., Klemola\,31B and LEDA\,2807038, have been estimated assuming a line width of 200\,\kms.


\begin{table*}
\centering
\caption{Properties of Klemola group members based on \hi\ 21-cm emission line}
\begin{tabular}{lcccccccc}

\hline\hline
&  &                                            ESO 400-11 &                                           ESO 400-13 &                                           Klemola\,31A &                              LEDA\,631562 &                               Klemola\,31B &                              LEDA\, 2807038 &                                        \\ 

\hline
(1) &R.A.                           &  $20^{{\rm h}}\, 23^{{\rm m}}\, 36.\!\!^\mathrm{s}96$ &  $20^{{\rm h}}\, 24^{{\rm m}}\, 0.\!\!^\mathrm{s}72$ &  $20^{{\rm h}}\, 23^{{\rm m}}\, 44.\!\!^\mathrm{s}88$ &  $20^{{\rm h}}\, 24^{{\rm m}}\, 6.\!\!^\mathrm{s}72$ &  $20^{{\rm h}}\, 23^{{\rm m}}\, 50.\!\!^\mathrm{s}16$ &  $20^{{\rm h}}\, 23^{{\rm m}}\, 34.\!\!^\mathrm{s}8$ \\ 
(2) &Declination                          &            $-36^{\circ}\,52\arcmin\, 51.\!\!\arcsec6$ &           $-36^{\circ}\,55\arcmin\, 55.\!\!\arcsec2$ &             $-36^{\circ}\,55\arcmin\, 4.\!\!\arcsec8$ &           $-36^{\circ}\,52\arcmin\, 26.\!\!\arcsec4$ &            $-36^{\circ}\,55\arcmin\, 19.\!\!\arcsec2$ &           $-36^{\circ}\,53\arcmin\, 16.\!\!\arcsec8$ \\           

(3) &$z_\hi$                            &                                             0.02867 &                                            0.02861 &                                             0.02872 &                                            0.02927 &                                                     - &                                                    - \\ 
(4) &$v_\hi$ (\kmps)                &                                                $8595$ &                                               $8577$ &                                                $8610$ &                                               $8776$ &                                                     - &                                                    - \\ 
(5) &S (Jy Km/s)                  &                                                $2.03\pm0.13$ &                                               $0.32\pm0.06$ &                                                $0.86\pm0.08$ &                                               $0.40\pm0.07$ &                                                     $<0.14$ &                                                   $<0.14$ \\ 
(6) &log \mhi\ (\Msun)            &                                       $9.86 \pm 0.03$ &                                      $9.05 \pm 0.09$ &                                        $9.50 \pm 0.04$ &                                       $9.20 \pm 0.08$ &                                                     $<$ 8.7 &                                                    $<$ 8.7 \\ 
(7) &MAJ\_s\                &                                              $38.7\arcsec$ &                                             $19.2\arcsec$ &                                              $32.7\arcsec$ &                                             $24.6\arcsec$ &                                                     - &                                                    - \\ 

(8) &MIN\_s\                &                                               $15.5\arcsec$ &                                             $11.8\arcsec$ &                                              $15.1\arcsec$ &                                              $18.7\arcsec$ &                                                     - &                                                    - \\ 
(9) &PA\_s\         &                                            $-47.1^{\circ}$ &                                           $-42.3^{\circ}$ &                                            $-65.9^{\circ}$ &                                           $-8.1^{\circ}$ &                                                     - &                                                    - \\ 
(10) &D$_{\hi}$ (kpc)              &                                               $44.7$ &                                               $23.4$ &                                               $38.0$ &                                              $28.8$ &                                                     - &                                                    - \\ 
(11) &D$_{\hi}/\rm D_{25}$ &  $1.04$ & $0.56$ &                                               $1.21$ &                                              $1.34$ &                                                     - &                                                    - \\ 

(12) &log $\rm{M^{exp}_{\hi}}$ (\Msun)           &                                       $9.98 \pm 0.21$ &                                      $9.89 \pm 0.21$ &                                       $9.72 \pm 0.21$ &                                      $9.28 \pm 0.21$ &                          $9.17 \pm 0.21$ &                                       $9.20\pm 0.21$ \\ 

(13) &DEF (\Msun)                  &                                               $0.12$ &                                              $0.84$ &                                               $0.22$ &                                              $0.08$ &  - & -                             \\

\hline
\end{tabular}

\begin{tablenotes}
\item

\label{tab:HI_properties}
\end{tablenotes}
\end{table*}


As previously discussed, \hi\ emission observations of Klemola\,31A have been reported in the literature. The total line flux of 0.78  Jy\,\kms\ reported by \citet{Boisse88} is consistent with our MeerKAT measurement of $0.86\pm0.08$\,Jy\,\kms. Klemola\,31A has the morphology of a barred `ocular' spiral galaxy exhibiting oval shaped bright structure at the center and a double spiral arm on the side of the galaxy away from Klemola\,31B \citep[see also Fig.~8 of][]{Carilli92}.  The formation of such structures is very short-lived (one rotation) and is sensitive to the strength of tidal interaction \citep[][]{Elmegreen91}.
The \HI\ disk also resembles the warped structure with large reservoir of neutral hydrogen gas trailing the far-side spiral arms from the galaxy B indicating a recent tidal disruption. We return to this point in Section~\ref{sec:hi_kinematics}.

Among all four galaxies detected in \hi\ emission, the most massive galaxy,
ESO\,400-11, is an edge-on spiral galaxy with total \HI\ flux of $2.03\pm0.13$ Jy \kms\ and \HI\ mass, \mhi\ $= 6.82\times10^9$\Msun.  As is usual for disk galaxies, the \hi\ disk extends well beyond the stellar disk.  It also shows a disturbed morphology, and similar to Klemola\,31A, has a gas poor (\mhi\ $<1.57\times10^7$\Msun) early-type galaxy (LEDA\,2807038) as a companion. 

The most remarkable contrast between the stellar and \hi\ components revealed by the MeerKAT imaging is observed for ESO 400-13, which is a S0 galaxy. 
The integrated \hi\ emission line flux and the corresponding \hi\ mass are $0.32\pm0.06$\,Jy\,\kms\ and $1.06\times10^9$\,\Msun, respectively. Interestingly, the entire \hi\ disk is systematically offset by $\sim$14\,kpc from stellar disk. 
Such segregation of \hi\ from the optical disc has been observed in a few other cases.  NGC4438 in the Virgo cluster core and  Arp\,105 are remarkable examples of displacement between \hi\ gas and optical disks \citep[][]{Combes88, Duc97}.

Lastly, the optically faint galaxy LEDA\,631562 has \HI\ flux of $0.40\pm0.07$ Jy \kms\ and the corresponding \HI\ mass of $1.41\times10^9$ \Msun. Curiously, the [N~{\sc ii}]/H$\alpha$ versus [O~{\sc iii}]/H$\beta$ ratios i.e., the Baldwin, Philips, \& Terlevich (BPT) diagnostic for this galaxy is consistent with ionization through a power-law spectrum, i.e., an active galactic nuclei \citep[][]{Chilingarian17}. The central AGN may also be responsible for the \hi\ hole observed at the center of this galaxy. 
In passing, we note that the BPT diagnostic is available for two other galaxies i.e., ESO\,400-11 and ESO\,400-13 from \textsc{RCSED2}\footnote{\href{https://rcsed2.voxastro.org/}{https://rcsed2.voxastro.org/}}, and are consistent with these being normal star-forming galaxies.  Only ESO\,400-11 is clearly detected\footnote{Excluding Klemola\,31A which can not be reliably investigated in radio continuum due to the proximity of PKS\,2020-370.} in radio continuum in the MALS wideband image (rms noise $\sim$30\,$\mu$Jy\,beam$^{-1}$) with a total flux density of 3.1\,mJy 
This implies a star formation rate (SFR) of 2.8\,\Msun\,yr$^{-1}$ \citep[][]{Bell03}. ESO\,400-13 is only tentatively detected in the radio continuum with a flux density of 0.4\,mJy (SFR $\sim$0.6\,\Msun\,yr$^{-1}$).

Overall, Klemola\,31 is a rare example of galaxy group at an early stage of its evolution with different morphologies and types of galaxies. Although we find recent gravitational disturbances in most of the individual galaxies, we do not detect any intra-group \HI\ emission in the form of bridges and clumps. 
Nevertheless, the \HI\ content and morphology of each galaxy in the group seems to be affected by the group environment or by interaction with other galaxies \citep[e.g., see][]{ Serra13, Serra15, Gupta18}. These processes may lead to galaxies becoming \hi\ deficient.   In order to obtain the \HI-deficiency we estimate the expected \HI\ masses for all the member galaxies of Klemola\,31 using the following scaling relation based on B-band luminosity \citep{Jones2018}: 
\begin{equation}
\label{eqn:expected_Hi_mass}
    \text{log}\, \rm{[M^{exp}_\hi/ M_{\odot}]} = X\, \text{log}\, [L_B\, / L_{\odot}] + \text{Y}
\end{equation}
 where, X and Y are gradient and intercept estimated by \cite{Jones2018} for different morphological type galaxies. The B-band luminosity ($L_B$) is derived by\\
\begin{equation}
\label{eqn:B-band_luminosity}
    \text{log}\, \rm{[L_B /L_{\odot}]} = 10 + 2\,\text{log}\, [D_L/\text{Mpc}] + 0.4 (M_{\text{bol},\odot} - B \rm ^c)
\end{equation}
where, $\rm{M_{\text{bol},\odot}} = 4.88$ is the bolometric absolute magnitude of the Sun \citep[][]{Lisenfeld11}, $\rm B \rm ^c$ is the corrected B-band magnitude after galactic extinction, internal extinction and k-correction.
The B-band magnitude for all the galaxies are obtained from HyperLeda\footnote{\href{http://leda.univ-lyon1.fr}{http://leda.univ-lyon1.fr}}.

\begin{figure}
 \includegraphics[width=\columnwidth]{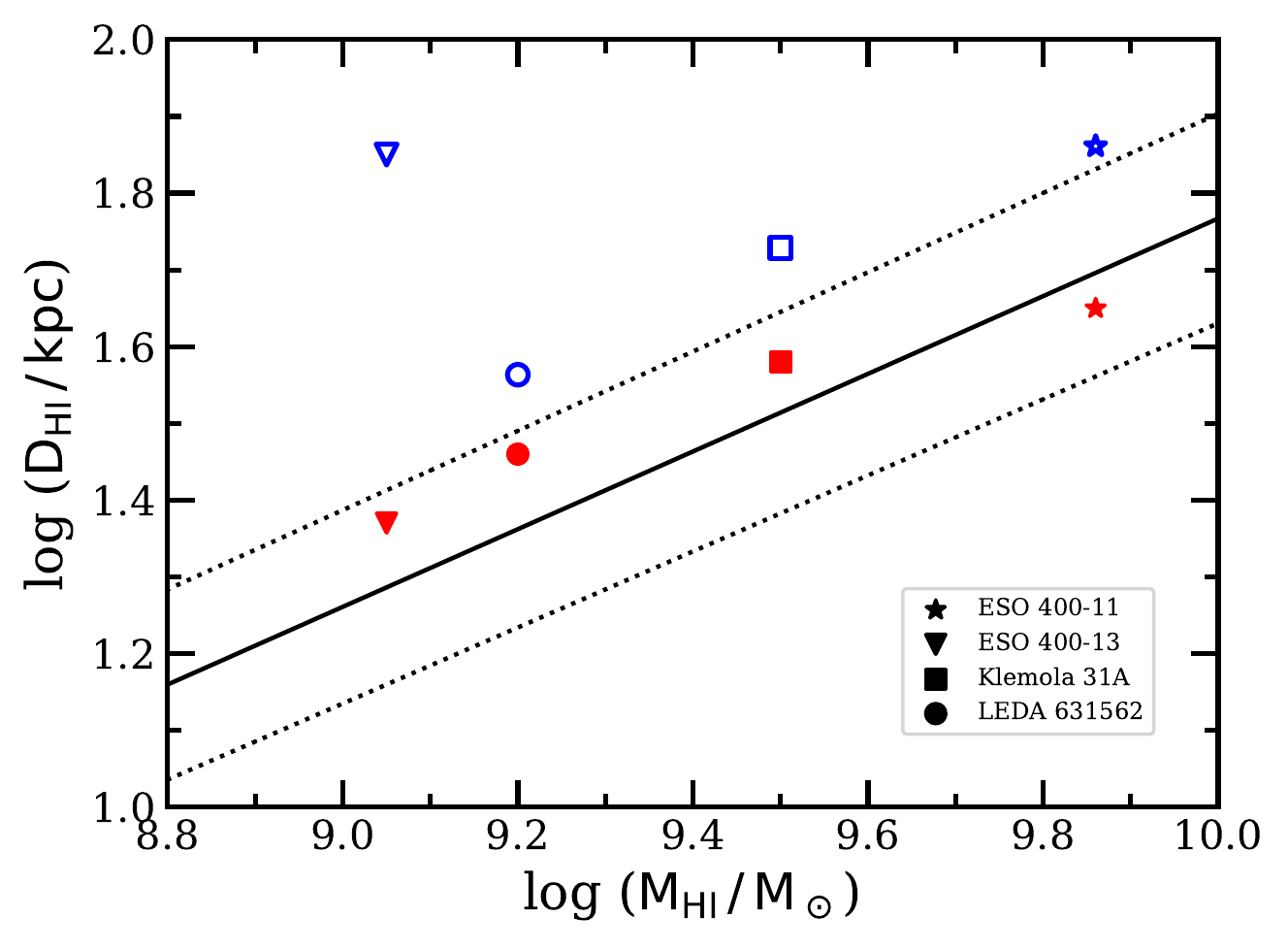}
 \caption{The $D_{\rm HI}-M_{\rm HI}$ relation for the detected galaxies.
 The red (filled) symbols are based on $D_{\rm HI}$ and $M_{\rm HI}$ based on our measurements.  The blue (open) symbols use $D_{\rm HI}$ predicted using the scaling relation based on optical diameter, $\rm D_{25}$ \citep[][]{Broeils97}.   Each galaxy has been shown using a different symbol.  The solid lines represent the best-fit linear relation from \citet{wang_new_2016} and the dashed line represents $3\sigma$ scatter.}
 \label{fig:size-mass}
\end{figure}


The expected \HI\ mass values (log $\rm{M^{exp}_\hi}$) for the galaxies calculated using  Eq.~\ref{eqn:expected_Hi_mass} are presented in Table~\ref{tab:HI_properties}.
The deficiency parameter, i.e., the deficiency between expected \HI\ mass and observed \HI\ mass is defined as, $\text{DEF$_{\rm{\ion{H}{I}}}$} = \text{log}\, \rm{M^{exp}_\hi/ M_{\odot}} - \text{log}\, M_{\ion{H}{i}}/ M_{\odot}$. A negative value of $\rm{DEF_{\hi}}$ parameter indicates \hi\ excess in the galaxy whereas a positive value suggests \hi\ deficiency. 
In the literature, to account for intrinsic scatter in the used scaling relations, the galaxies are usually considered to be \hi\ deficient when $\rm{DEF_{\hi}} >$0.2-0.3\,dex.  Considering this, only ESO\,400-13 shows significant \hi\ deficiency ($\rm{DEF_{\hi}}$ = 0.84). 
This is not surprising as this galaxy also shows the maximum displacement between the optical and \hi\ disks (Fig.~\ref{fig:Mom0_col_dens}). The \hi\ gas is totally asymmetric, being pushed on one side of the galaxy only. This is characteristic of a ram-pressure stripped galaxy, running at high velocity in a hot and tenuous intra-galactic gas (IGM).

Examples of HI detected in only one side of galaxies can be found in the Coma cluster \citep{chen2020}. It is however unique in galaxy groups. Klemola\,31 is a small group, and is not expected to contain a dense IGM, thus the ESO\,400-13 galaxy must be crossing the group at very high velocity, mainly in the plane of the sky, since its redshift is comparable to the other group galaxies. 
For stripping to occur, the ram pressure on the \hi\ needs to be greater than the restoring gravitational force. The ram-pressure ($P_{ram}$) as well as the anchoring self-gravity $\Pi$ from ESO\,400-13, can be modelled following \citet[][]{Gunn72} as
\begin{equation}
    P_{ram} = \rho_{IGM}v^2_{gal}, {\rm and}
\label{eq:gunn1}    
\end{equation}
\begin{equation}
    \Pi = 2\pi\mathrm{G} \Sigma_s \Sigma_g
\label{eq:gunn2}
\end{equation}
Here, $\Sigma_s$ and $\Sigma_g$ are stellar and gas surface densities, respectively.
The B-band magnitude and $D_{25}$ of ESO\,400-13 (Table~\ref{tab:group_gal}), for a mass-to-light ratio of unity correspond to $\Sigma_s$ $\approx$ 16\,$M_\odot$\,pc$^{-2}$. 
Similarly, assuming that the galaxy prior to stripping had ${\rm M^{exp}_\hi}$ distributed over $D_{25}$, we get  $\Sigma_g$ $\approx$ 8\,$M_\odot$\,pc$^{-2}$.  If the gas is distributed over a larger area, as is typically the case, then it will be even more susceptible to stripping.  For a typical density of hot gas (3$\times10^{-3}$\,cm$^{-3}$) in the center of the group, the equations~\ref{eq:gunn1} and \ref{eq:gunn2} imply a typical relative velocity of $\sim$200\,\kms.  This is totally feasible, and compatible with the hypothesis of ram pressure stripping for ESO\,400-13.

\begin{figure}
\centering
 \includegraphics[scale=.5]{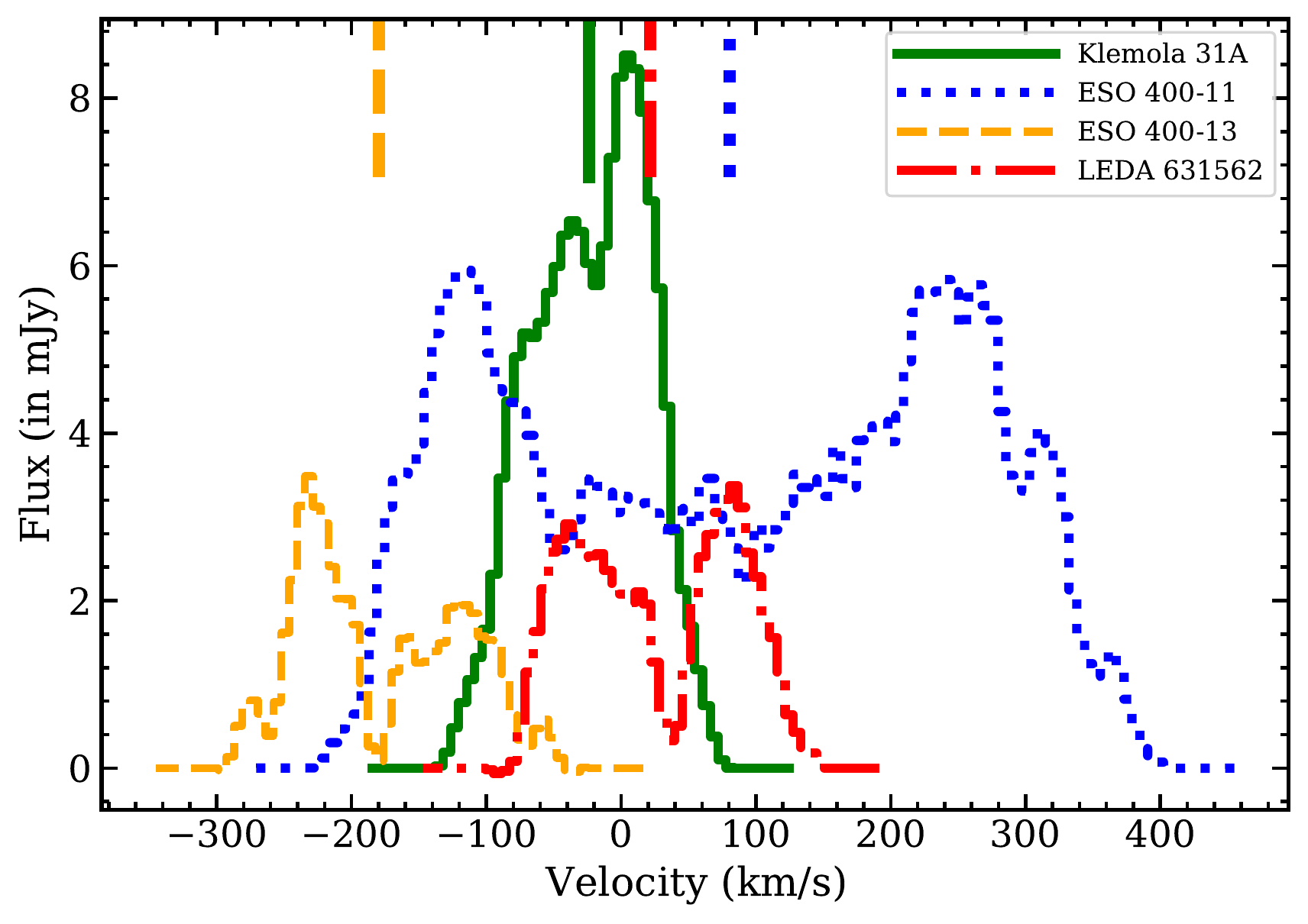}
 \caption{The global \HI\ profiles of Klemola\,31A, ESO 400-11, ESO 400-13 and LEDA 631562. The zero of the velocity scale is defined with respect to optical redshift, $z_{g}$ (see Table~\ref{tab:group_gal}). The dashed vertical lines denote $z_\hi$.}
 \label{fig:spectra}
\end{figure}

We estimated the \HI\ diameter ($\rm D_{\ion{H}{I}}$) of each galaxy by fitting ellipses to the \HI\ column density contour map corresponding to 1\,[$\rm{M_{\odot}pc^{-2}}$] = $1.25\times10^{20} \rm{cm^{-2}}$.
For Klemola\,31A, the estimated $\rm D_{\ion{H}{I}}=38.02$ kpc is consistent with the value provided by \citet{Carilli92}. ESO 400-11 has the largest \HI\ disk diameter of 44.7 kpc in the group, whereas, ESO 400-13 and LEDA 631562 have $\rm D_{\ion{H}{I}}$ of only 23.4 kpc and 28.8 kpc, respectively.
In Fig.~\ref{fig:size-mass}, red (filled) symbols represent  $\rm D_{\rm HI}$ and $\rm M_{\rm HI}$ based on our measurements.
The solid lines represent the best-fit linear relation from \citet{wang_new_2016} and the dashed line represents $3\sigma$ scatter.  Clearly, all group members detected in \HI\ are in compliance with the relation. Only the expected \hi-size of ESO 400-13 is strikingly larger than its measured size, and this supports the ram-pressure stripped character of this galaxy.

\begin{figure*}
\centering
 \includegraphics[scale=.55]{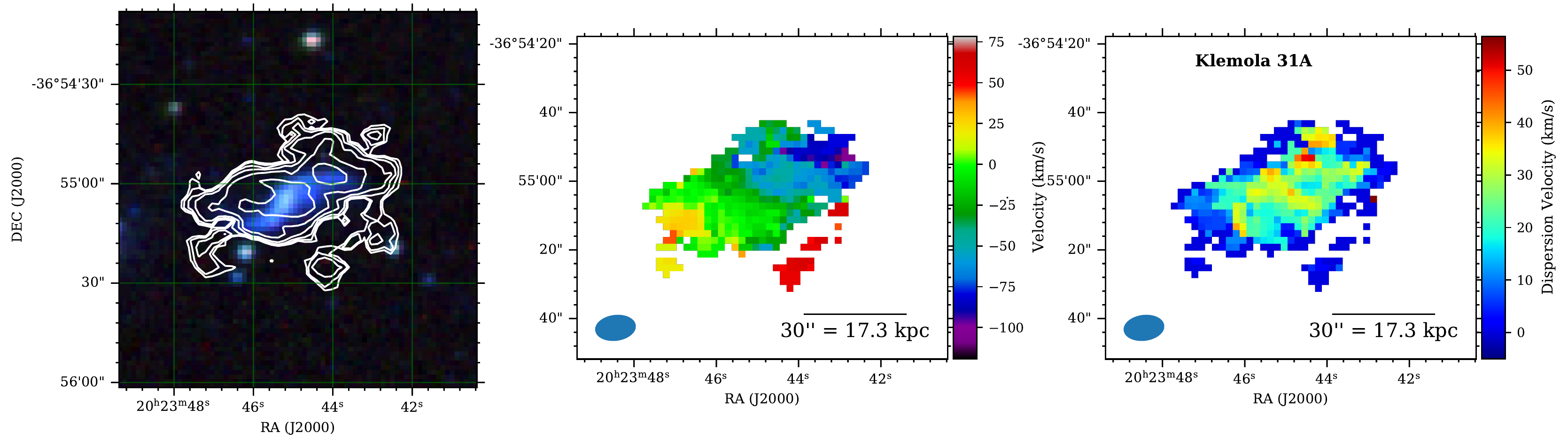}\\
  \includegraphics[scale=.55]{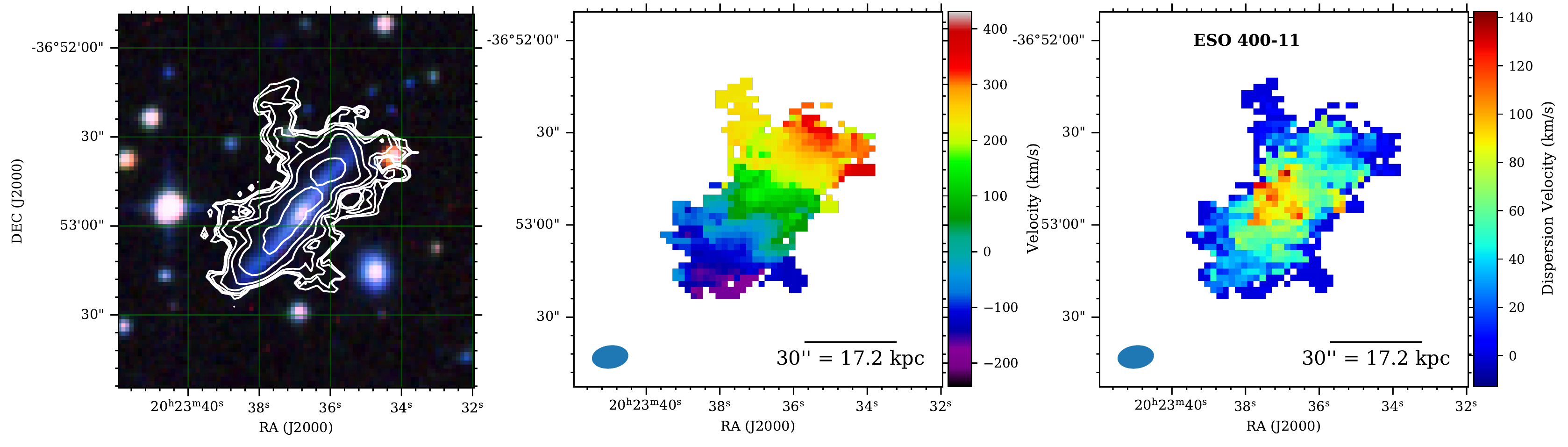}\\
   \includegraphics[scale=.55]{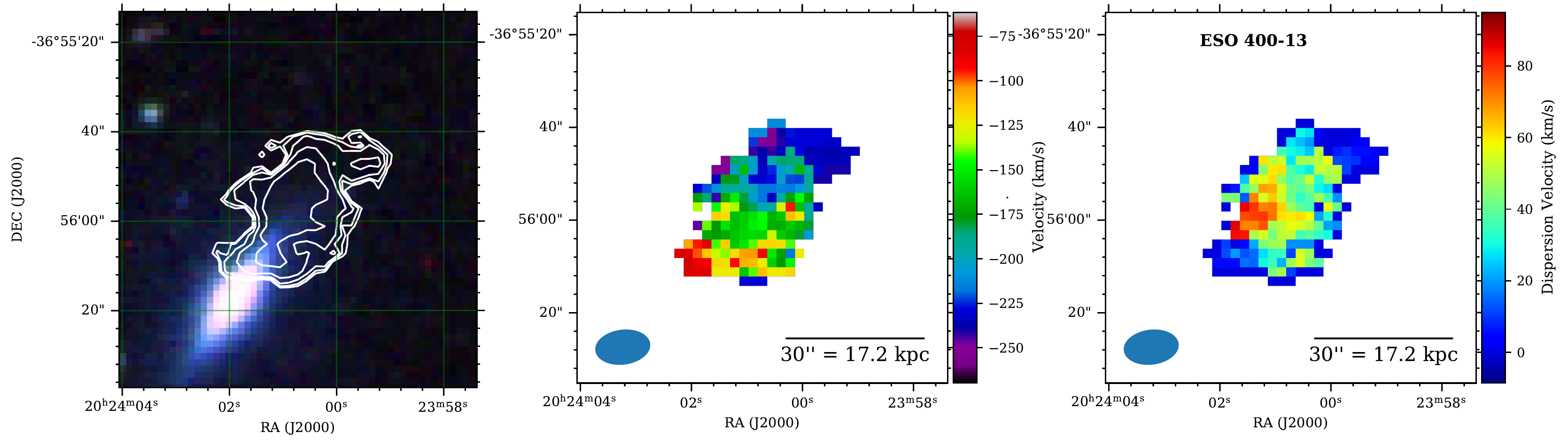}\\
    \includegraphics[scale=.55]{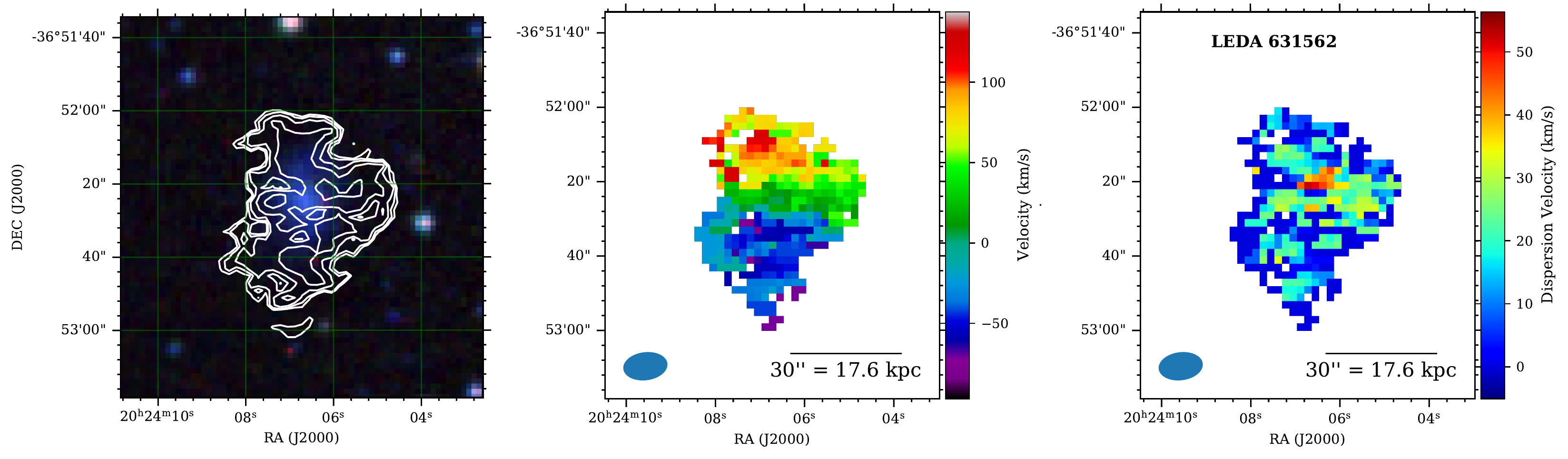}\\
 \caption{\HI\ column density (moment-0) contours plotted over RGB cutouts from the optical DSS image (left), moment-1 maps (middle) and moment-2 maps (right) of four detected \HI\ sources, Klemola\,31A, ESO 400-11, ESO 400-13, and LEDA 631562. The zero of the velocity scales in middle panels are defined with respect to $z_{g}$ based on the optical spectra (Table~\ref{tab:group_gal}). The contours in the middle panels correspond to \HI\ column densities of $ 2^n \times 5 \times 10^{19}$ \,cm$^{-2}$ ($n=0,\,1,\,2,\,3,\, ...)$. 
 The MeerKat restoring beam of $11\farcs9\arcsec \times 7\farcs6\arcsec$ (6.9 kpc $\times$ 4.4 kpc) and a spatial scale bar of $30\arcsec$ are also shown in middle and right panels.}
 \label{fig:combined_image}
\end{figure*}

\hi\ disks of galaxies are known to extend well beyond the optical disks by about a factor of $\sim$2 \citep[][]{Broeils97}.  The optical diameters i.e., $\rm D_{25}$ of the group members are provided in in Table~\ref{tab:group_gal}. We find the \hi\ disks of Klemola group members to be significantly truncated with respect to their optical disks (see row 11 of Table~\ref{tab:HI_properties}).  If the $D_{\rm HI}$ of these galaxies were truly representative of the $D_{\rm HI}$-$\rm D_{25}$ scaling relation, log $\rm D_{\hi} = (1\pm0.03)\, log\, D_{25} + (0.23\pm0.04)$, from  \citet[][]{Broeils97}, the corresponding points would lie well above the $D_{\rm HI}-M_{\rm HI}$ relation (shown with blue/open symbols in Fig.~\ref{fig:size-mass}) 
Unsurprisingly, the most significant deviation is observed for ESO\,400-13, an early-type galaxy with $\rm D_{\hi}/D_{25}=0.56$. 
Similar deviations are also observed in the case of Virgo cluster spirals \citet[][]{Cayatte94}.
\citet[][]{reynolds_wallaby_2021} have also recently observed truncation of the \hi\ disks of Hydra cluster galaxies compared to their field galaxy counterparts, and attribute this to earlier stages of losing \hi\ gas reservoirs likely as a  result of environmental processes stripping the gas (e.g. ram pressure and tidal stripping).
All this demands further investigation of gas removal processes in different environments which affect the $\rm D_{\hi}-D_{25}$ relation, but leave the $\rm D_{\rm HI}-\rm M_{\rm HI}$ relation remarkably intact \citep[][]{Stevens19}.

\subsection{Galaxy group \hi\ kinematics}
\label{sec:hi_kinematics}

The high spatial resolution capabilities of the MeerKAT allows for kinematics analysis of the group. Despite the disturbed \hi\ morphology, all the galaxies are resolved above 3-beam elements along the projected major axis.
Fig. \ref{fig:spectra} displays the global spectra of the four detected galaxies. The \hi\ spectra have been centered on their optically-defined redshifts. Note that ESO\,400-13 is detected only on the blue-shifted side. Klemola\,31A and ESO 400-13 show kinematic asymmetry in comparison to the other two galaxies. The lopsided gas distribution of Klemola\,31A and ESO 400-13 are clearly visible in Fig.~\ref{fig:spectra}. 
Such asymmetries can arise mainly due to three different environmental mechanisms, i.e., (1) galactic interaction, (2) gas accretion from neighbouring filaments and (3) ram-pressure stripping  \citep[e.g.,][]{Chung07, Mapeli08, Blok14, Denes16, Elagali19}. Our observations do not reveal any prominent \hi\ bridge/tail that would arise from tidal interactions or minor mergers with neighbour galaxies. However, clear signature of ram pressure striping are evident for ESO 400-13, which shows significant compressed \hi\ gas (by factor of $\sim2$) in the receding velocity side towards Klemola\,31A. 

The first three moments of the individual \hi\ cubes, i.e., surface density, velocity field, and velocity dispersion, are plotted in Fig. \ref{fig:combined_image}. The velocity fields for three galaxies are only slightly perturbed by interactions, while the ram-pressure stripped ESO\,400-13 has the most irregular spectrum and velocity field. Klemola 31A reveals a characteristic distortion in its SE extension, due to the competition between emission and absorption in front of PKS\,2020-370. There is also a SW tidal extension, displaying receding velocities, suggesting that the tail is not in the same plane as the galaxy, dragged out by the interaction with Klemola 31B. ESO\,400-11 reveals also tidal extensions, due to the action of LEDA 2807038. LEDA 631562 has the most regular velocity field, despite the central \hi\ hole.  ESO\,400-13 reveals the characteristic perturbation of ram-pressure stripping, suggesting a high velocity with respect to the rest of the group, mainly in the plane of the sky. Only half of the \hi\ velocity field is observed.

The velocity dispersions are in concordance with the findings of the velocity maps. There are only small perturbations with respect to the usual feature, of a maximum dispersion along the minor axis of galaxies, due to the strong gradient in this region of the spider diagram. Only for the ram-pressure stripped galaxy, ESO\,400-13, where only one side is \hi\ detected, and the minor axis is deficient, the dispersion map is unusual and strongly perturbed.

\subsection{Line of sight absorption towards PKS\,2020-370}
\label{sec:hiabs}

\begin{figure} 
\centerline{
\includegraphics[trim= {0cm 5cm 0cm 0cm}, width=0.55\textwidth,angle=0]{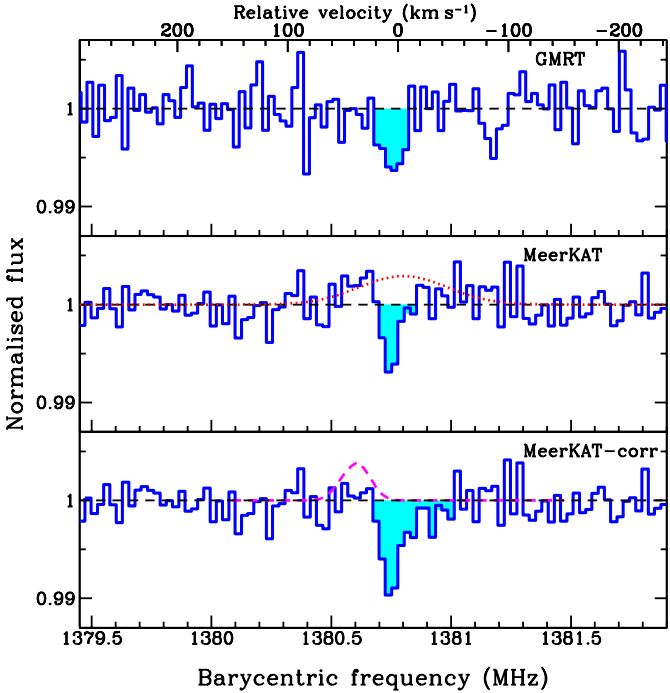}  
}  
\caption{\hi\ 21-cm absorption towards PKS\,2020-370 (spectral resolution $\sim$5.7\,\kms).  
The zero of the velocity scale is defined with respect to the peak of the absorption i.e., $z_{21cm}$ = 0.0287.  The dashed line in the middle panel represents the model for the \hi\ emission.  The bottom panel shows absorption profile corrected using this model (see text for details). 
Overplotted in magenta is a predicted emission line profile, based on the model of a regularly rotating disk. The model has been constructed by fitting a tilted-ring model to the data cube, as described in Sect.~\ref{sec:hiabs}. 
} 
\label{fig:hiabs}   
\end{figure} 

The \hi\ 21-cm absorption spectra towards the quasar are shown in Fig.~\ref{fig:hiabs}. The GMRT spectrum presented in the top panel has been smoothed by 3 channels to obtain a spectral resolution comparable to the MeerKAT spectrum ($\sim$5.7\,\kms).  The absorption signal in the MeerKAT spectrum is clearly contaminated by the \hi\ emission (see middle panel of Fig.~\ref{fig:hiabs}). This emission is much less evident in the GMRT spectrum but it is clear that the continuum subtraction is not perfect here as well.  Simply using the frequency ranges marked by the shaded regions in the top and middle panels, we measure $\int\tau dv$ = $0.13\pm0.04$\,\kms\ (GMRT) and $0.09\pm0.02$\,\kms\ (MeerKAT), respectively.  In both the cases, the unaccounted systematic error due to continuum subtraction dominates the optical depth estimate.

We focus on the more sensitive MeerKAT spectrum and correct the absorption profile for contamination from emission as following. First, we mask the frequency channels detected in absorption, shown as shaded region in the middle panel, in the MeerKAT spectrum. Next, we model the emission in the remaining channels using a single Gaussian component (see dotted line in the middle panel). The absorption profile corrected using this model is shown in the bottom panel of Fig.~\ref{fig:hiabs} and corresponds to $\int\tau dv$ = $0.26\pm0.03$\,\kms. 
Still, the uncertainties in determining the emission profile to use for the continuum subtraction would dominate. Nevertheless, the estimated integrated absorption matches well within 1$\sigma$ with the VLA A-configuration measurement by \citet[][]{Carilli87}.  Later in the section, we attempt to address the limitations of this simplistic approach.

For optically thin gas, the \hi\ 21-cm optical depth is related to the \hi\, column density, covering factor $f_c$ and spin temperature $T_{\rm s}$ through 
\begin{equation}\label{eqn:absorption_col_density}
    \frac{N(\ion{H}{i})}{\mathrm{cm^{-2}}} = 1.823 \times 10^{18} \frac{T_s}{\mathrm{K}}\frac{1}{f_c} \int \tau(v) \frac{dv}{\mathrm{km\,s}^{-1}}
\end{equation}
PKS\,2020-370 has a flat spectral index, $\alpha\approx 0.2$, over 0.8 - 20\,GHz, and based on WISE mid-infrared colors can be classified as a blazar \citep[][]{Dabrusco19, Gupta21}.  Both of these suggest that the radio emission is core-dominated.  The milliarcsecond scale images at 2.3\,GHz recover 85\% of the total flux density \citep[][]{Petrov17}, of which 90\% is in the central `core-jet' component with an extent of $0\farcs015$ (9\,pc at $z_{grp}$ = 0.029).  
The typical sizes of CNM clouds producing \hi\ 21-cm absorption is $>5$\,pc and they may be part of cold gas structures that extend beyond $\sim$35\,pc \citep[][]{Gupta18}.  
Further, constraints on the covering factor can only be obtained through milliarcsecond scale resolution spectroscopy \citep[see also][]{Srianand13dib}. 
Here, we proceed with the reasonable assumption of $f_c = 1.0$, and estimate $N$(\hi) = (4.7$\pm$0.6)$\times10^{19}({T_s \over 100\,\mathrm{K}})({1.0 \over f_c})\,\mathrm{cm}^{-2}$.
The \hi\ emission model used to correct the MeerKAT absorption profile corresponds to $N$(\hi) = $1.4\times10^{21}$\,\cmsq.  For $\int\tau dv$ = $0.26\pm0.03$\,\kms, this corresponds to the limit, $T_{\rm s} <$ 2950\,K.  Naturally, this is a conservative upper limit since \hi\ emission is integrated over an area of 6.9 kpc$\times$4.4 kpc, and cannot be compared with absorption if there are inhomogeneities and clumpiness.

\begin{figure*}
\centering
\includegraphics[width=0.98\textwidth]{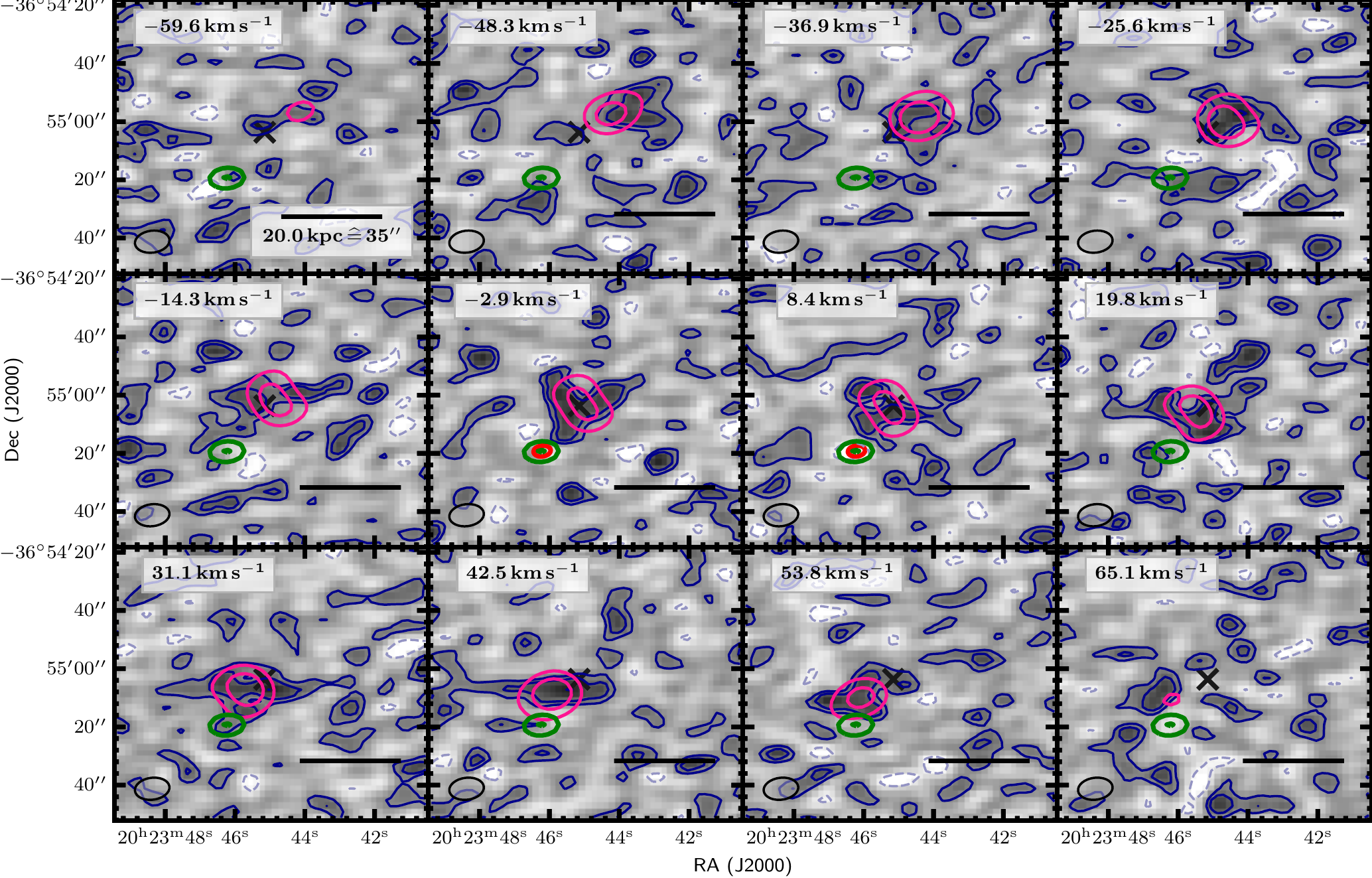}
\caption{\ion{H}{i} data cube resulting from the MeerKAT observations (blue contours and greyscale), {\tt{TiRiFiC}} model data cube (pink contours), and continuum image of PKS~2020-370 (green contours). Every second channel of the data cube is shown. The velocity is measured relative to $z\,=\,0.0287$ at the peak velocity of the absorption feature against PKS~2020-370. The red ellipses mark channels in the velocity range of the absorption feature ($FWHM\,\approx\,16\,\mathrm{km}\,\mathrm{s}^{-1}$). Dark grey cross: centre of {\tt{TiRiFiC}} model. Contours corresponding to line emission scale by $\sigma_\mathrm{rms}\,=\,0.79\,\mathrm{mJy\,beam}^{-1}$. Light blue and dashed contours (MeerKAT data cube): $-2\,\sigma_\mathrm{rms}$. Blue and pink contours (MeerKAT data cube and {\tt{TiRiFiC}} model): $1,\,2,\,4\,\sigma_\mathrm{rms}$. Green contours (radio continuum): $200,\,400\,\mathrm{mJy\,beam}^{-1}$. The black ellipse in the lower left corners of the channel maps denotes the synthesized beam at half power.}
\label{fig:cube}
\end{figure*}



\begin{figure}
\centering
\includegraphics[width=0.48\textwidth]{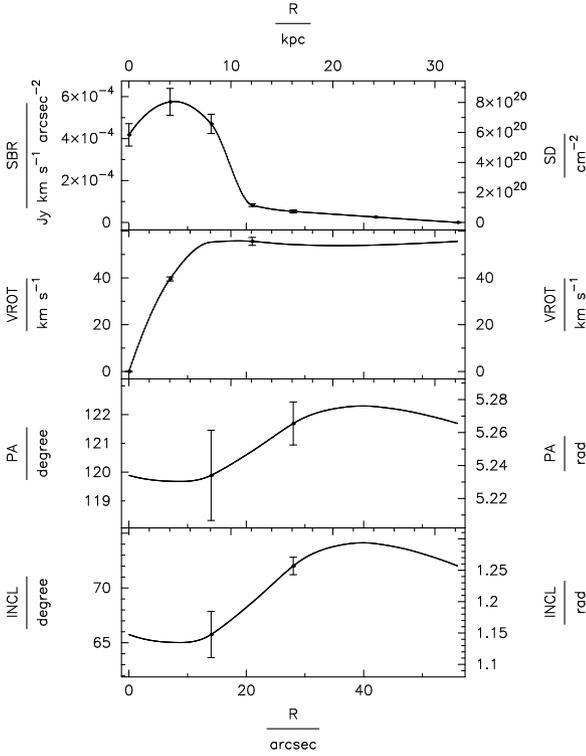}
\vspace{-1cm}
\caption{{\tt{TiRiFiC}} tilted-ring model of the \ion{H}{i} disk of Klemola 31A. The curves are spline fits, the edge points having the same value as the last data point for each parameter. SBR: surface brightness. SD: surface column density. VROT: rotation velocity. PA: position angle. INCL: inclination.}
\label{fig:tirmodel}
\end{figure}

\begin{figure}
\centering
\includegraphics[width=0.48\textwidth]{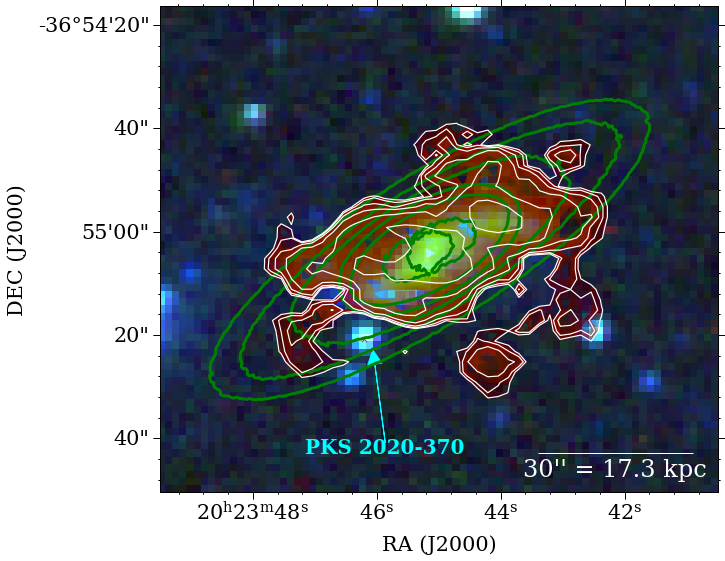}
\caption{Observed moment-0 map of Klemola\,31A in autumn color gradient (increasing $N(\hi)$ from brown to yellow) overlaid on the RGB cutout of DSS image. We also show the observed (white) and {\tt{TiRiFiC}} tilted-ring model (green) moment-0 contours with the same levels as Fig.~\ref{fig:combined_image}.}
\label{fig:Mom1}
\end{figure}

To estimate the \ion{H}{i} column density, and derive the spin temperature of the \ion{H}{i}, at the position of PKS~2020-370 in an alternative approach, we created a tilted-ring model, using a direct fit to the data cube. 
Such a model will also allow us to probe the  line of sight velocity field probed by absorption in the framework of rotating HI disk.  For this, we used a cutout of our data cube around the position of Klemola~31A, which is shown in Fig.~\ref{fig:cube}. We then applied the tilted-ring modelling software FAT \citep{Kamphuis2015b} to get an initial tilted-ring model. In the following, {\tt{TiRiFiC}} \citep{Jozsa2007}, a computer program, which also optimises tilted-ring parameters through simulating spectroscopic 3-D data cubes of rotating galactic disks, was used to simplify this model. While FAT is a fully automated software and uses {\tt{TiRiFiC}} as the underlying engine, {\tt{TiRiFiC}} in the standalone mode allows for somewhat higher flexibility, in particular the sampling of the parameters at different intervals, a feature, which has been made use of in this study. The tilted-ring parameters surface brightness, rotation velocity, inclination and position angle were distributed on nodes at radii between $R\,=\,$0\carcsec\ and $R\,=\,56^{\prime\prime}$ and automatically optimized. We use an Akima interpolation between the single nodes, while nodes at the edge of the disk (at radii $R\,=\,0^{\prime\prime}$ and $R\,=\,56^{\prime\prime}$) were chosen to assume the value of the closest node for the rotation velocity, the position angle, and the inclination. This implies that the rotation curve is approximately flat. The surface brightness was chosen to taper off to a value of 0 at the outermost radius $R\,=\,56^{\prime\prime}$. The centre of the model, the systemic velocity, and the dispersion were fitted to be constant with radius. After attempting several combinations of parameters, and calculating errors as described in \citet[][]{Jozsa2021a}, we arrive at a slightly warped disk with a parametrization as shown in Fig.~\ref{fig:tirmodel}. Fig.~\ref{fig:cube} shows model data cube overplotted on the original data.  While the signal-to-noise ratio in the data cube is relatively high, owing to the high spectral resolution of the cube, we get a reasonable, albeit not perfect, match with the data. Fig.~\ref{fig:Mom1} shows the column density map of the original data and the model. Here also, it is evident that the data are not reproduced perfectly. Accepting the corresponding uncertainties, we obtain an \ion{H}{i} emission line model at the position of PKS 2020-370 at our hands, which we next compare to the observed absorption profile.

To predict the column density and the velocity of \hi\ gas at the location of PKS\,2020-370 from our model without being hampered by beam smearing effects, we created a model data cube at a very high spatial resolution (pixel size of $0.2^{\prime\prime}$ and a half-power-beam-width of $0.6^{\prime\prime}$).  Next we created a column-density map and obtain the column density of $N$(\hi) = $2.5\times10^{20}$\,\cmsq\ at the location of the PKS\,2020-370. For $N$(\hi) = $2.5\times10^{20}$\,\cmsq\ and $\int\tau$dv = 0.26\,\kms, 
we can then calculate the corresponding $T_{\rm s}$ = 530\,K for a homogeneous regularly rotating disk. This result relies solely  on the assumption that the morphology is reflected by our tilted-ring model, and does not take the kinematics into account.

Next, we make use of the kinematical model to estimate whether the absorption profile indeed arises from a regularly rotating disk. The predicted corresponding \hi\ emission profile is shown in the bottom panel of Fig.~\ref{fig:hiabs}. The emission peak, corresponding to $N$(\hi) = $2.5\times10^{20}$\,\cmsq, has a dispersion of 12.9\,\kms\ and is offset with respect to absorption peak by 38.0\,\kms\ ($z_{21cm}$=0.0287). The observed absorption profile hence stems from gas, which is not participating in the (assumed) regular rotation of the disk, it is anomalous gas. This can also be seen directly from the data cube (Fig.~\ref{fig:cube}). At the velocity of the absorption feature against PKS~2020-370, the peak of the emission has a significant offset from the position of the background source (Fig.~\ref{fig:cube}, central panels). The absorption line feature has hence a velocity corresponding to the systemic velocity, but not corresponding to the receding (South-East) portion of the disk, which would be expected according to our simple model: the absorption would occur at the edge of the receding side of the disk (Fig.~\ref{fig:cube}, central bottom panels).

\begin{figure}
\centering
\includegraphics[viewport=30 163 600 700,clip=true,width=0.5\textwidth]{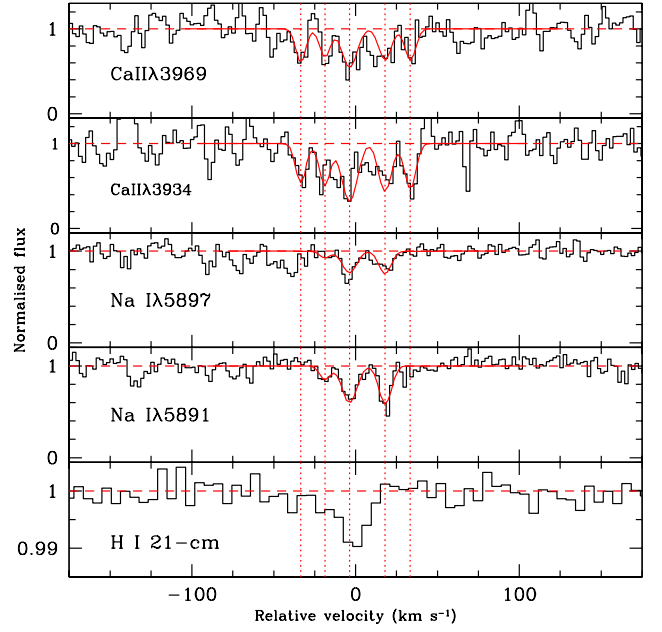}
\caption{Comparison of the absorption profiles \caii, \nai\ and \hi\ 21-cm absorption associated with Klemola\,31A towards PKS\,2020-370. The best fitted Voigt profiles (components identified with red vertical dotted lines) are over-plotted.  The zero velocity scale is defined at \zabs = 0.028725.
}
\label{fig:metal_absorption}
\end{figure}

To further explore the gas distribution along the line of sight,
we obtained the Keck/HIRES (High Resolution Echelle Spectrometer) raw data of 
PKS\,2020-370 from Keck Observatory Archive (KOA)\footnote{\href{https://koa.ipac.caltech.edu/}{https://koa.ipac.caltech.edu/}}.
We processed the raw data using {{\tt IRAF}} to obtain the continuum normalized spectrum, and identified the \caii\ and \nai\ metal lines associated with Klemola\,31A \citep[see also][]{Junkkarinen94}. 
We fitted Voigt profiles to the metal absorption lines using {{\tt VPFIT}}\footnote{\href{http://www.ast.cam.ac.uk/~rfc/vpfit.html}{http://www.ast.cam.ac.uk/~rfc/vpfit.html}}. In this case we assume the b-parameter to be same for \nai\ and \caii.
We show the absorption profile and best-fitted Voigt profiles in Fig.~\ref{fig:metal_absorption}, and present the Doppler parameter ($b$) and the column density of the fitted components in Table~\ref{tab:voigtfit}. From Fig.~\ref{fig:metal_absorption}, we can see the \caii\ absorption is spread over $\sim$ 80 \kms (between -40 to +40 \kms with respect to \zabs = 0.028725) in 5 distinct components. \nai\ absorption is clearly detected in the three central components two of them coinciding with the \hi\ 21-cm absorption. 
The measured velocity spread clearly supports the presence of additional gas components (with anomalous velocities) compared to the \hi\ disk.

The observed \nai/\caii\ ratios are consistent with what is typically  seen in gas producing 21-cm absorption in quasar-galaxy pairs where the line of sight does not pass through the gaseous or stellar disk \citep[][]{Dutta17}.  In the Milky Way,  we typically observe $N$(\nai)$>N$(\caii) \citep[see Fig.~9 of ][]{Welty96}.  Thus, it appears that the Ca depletion is unlike 
what we see in the cold ISM (i.e., about 3 orders of magnitude).  Since \nai\ is not shielded to ionizing radiation it is possible that the low \nai/\caii\ ratio towards PKS2020-370 reflects low radiation field as compared to the Galactic background radiation \citep[][]{Dutta17}.
Thus, based on kinematics and the line ratio discussed above it appears most of the absorption is produced by gas unrelated to cold \hi\ disk.

%
\begin{table}

    \centering
\caption{Metal absorption components of Klemola\,31A toward PKS\,2020-370.}
\small{
\begin{tabular}{ccccc}
\hline
\hline
$z_{abs}$ & $b$ & $ \rm log\, N(\rm \ion{Ca}{II})$  &   $ \rm log\, N(\rm \ion{Na}{I})$  & Notes
 \\
 & (in \kms) & (in $\rm cm^{-2}$) & (in $\rm cm^{-2}$) & \\
(1) & (2) & (3) & (4) & (5) \\
\hline \\

0.028661 & $2.02\pm1.77$ & $12.08\pm0.26$ & $11.12\pm0.11$ & Tied-$b$\\ 
0.028713 & $4.47\pm0.70$ & $12.32\pm0.09$ & $11.76\pm0.04$ & Tied-$b$ \\ 
0.028787 & $3.21\pm0.70$ & $12.15\pm0.11$ & $11.74\pm0.04$ & Tied-$b$ \\ 
0.028611 & $1.12\pm1.10$ & $12.44\pm1.25$ & $-$ \\ 
0.028840 & $2.16\pm1.73$ & $12.14\pm0.26$ & $-$ \\ 
\hline
\end{tabular}
}
\label{tab:voigtfit} 
\end{table}

There are two possibilities to explain the anomalous velocity of the absorption feature. Firstly, it might come from anomalous gas in the disk. Streaming motions due to the spiral arms, and velocity perturbations due to tidal interaction could cause the offset. In this case, the column-density estimate from the tilted-ring model and with that the estimate of $T_\mathrm{s}\,=\,530\,\mathrm{K}$ might still be valid. However, given a maximal rotation velocity of $56\,\mathrm{km\,s}^{-1}$ and an estimated projected rotation velocity of $53\,\mathrm{km\,s}^{-1}$ (for an inclination of $72\degree$), the offset of $38\,\mathrm{km\,s}^{-1}$ is significant, and the occurrence of such strong streaming motions along spiral arms for lower-mass galaxies might be more difficult to justify. Alternatively, the gas seen in absorption might not come from the disk. 

Overall, our observations may also indicate an extraplanar origin of the gas. In this case the spin temperature would remain an unknown, as we would not have a good measure of the \ion{H}{i} column density in emission. The \hi\ tidal extension to Klemola 31A towards the SW (see Sect.~\ref{sec:hi_kinematics}) might provide an explanation of the occurrence of the extraplanar gas in the sight line of the quasar. Assuming a tidal stream of \hi\ debris occurring between Klemola 31A and Klemola 31B, with the observed tidal extension to Klemola 31A being the sole visible part, any part of a hypothetical stream should have a velocity between the systemic velocity of Klemola 31B and the tidal feature. Klemola 31B is blueshifted with respect to the absorption line and the observed tidal feature redshifted. Our observations hence would not stand in contrast to the hypothesis of the origin of the absorption feature being tidal debris from an interaction between Klemola 31A and Klemola 31B.
As our estimate of $T_\mathrm{s}$ is based on a model of a homogeneous, circularly symmetric disk, and we know that the galaxy group, and with it, Klemola\,31A undergoes tidal interactions, we can only take these results as an indication.
An independent estimate of $N$(\hi) through ultraviolet spectroscopy  will help to derive the harmonic mean temperature along the line-of-sight \citep[e.g.,][]{Keeney05}.

\section{Summary and Conclusions}
\label{sec:summ}

The wide field of view and large instantaneous bandwidth of the MeerKAT array permits the simultaneous completion of several investigations requiring sensitive \hi\ 21-cm emission and absorption, and radio continuum imaging. Our aim was to identify with high spatial resolution the interstellar medium responsible for the \hi\ absorption in front of the quasar PKS\,2020-370.  Located in the outer parts of the \hi\ disk of Klemola\,31A, the line of sight reveals both emission and absorption, perturbing significantly the derived \hi\ surface density and velocity field of the foreground galaxy.

Out of the six identified galaxies of the Klemola 31 group, in the MeerKAT field of view, four are detected and well resolved in \hi. Three of them are only slightly perturbed, two through tidal interactions with small companions. They reveal tidal extensions, and out of the plane velocities, but not significant gas depletion. The fourth one, ESO\,400-13 is strongly \hi-deficient, and reveals a totally asymmetric \hi\ distribution, characteristic of ram-pressure stripped galaxies. The \hi\ is detected only on the approaching side of the galaxy. This is suggesting that the galaxy is crossing the group with a high velocity, mainly in the plane of the sky.

Using symmetry arguments and tilted-ring modelling of Klemola\,31A, we derive for the gas absorbing in front of PKS2020-370 a spin temperature of $T_\mathrm{s}\,=\,530\,\mathrm{K}$ under the assumption that the absorption arises from a symmetric disk. This rather high value, unexpected for a disk, as well as the velocity offset of the absorption feature from the expected radial velocity of the disk model, might indicate an extraplanar origin of the absorption feature. One explanation could be that we observe tidal debris, as the galaxy group is in a rather dynamical state, as indicated by asymmetric kinematics and morphology of the group members. The same interaction, however, might be responsible for a deviation of the kinematics of Klemola 31A from the assumed circularity of orbits, in which case the discrepancy of the velocities might be due to streaming motions.

This work demonstrates that the MALS large program, dedicated to absorption studies, will lead to environmental studies in galaxy groups through their \hi\ emission mapping, and help in understanding the origins of gas responsible for \hi\ 21-cm absorption.

\section*{Acknowledgements}
\label{sec:ack}

We thank MeerKAT and GMRT staff for their support during the observations. We also thank the referee for helpful comments. The  MeerKAT  telescope  is  operated  by  the  South  African  Radio  Astronomy  Observatory,  which  is  a  facility of   the   National   Research   Foundation,   an   agency   of   the   Department   of   Science   and Innovation. GMRT is run by the National Centre for Radio Astrophysics of the Tata Institute of Fundamental Research.  
This research has made use of the Keck Observatory Archive (KOA), which is operated by the W. M. Keck Observatory and the NASA Exoplanet Science Institute (NExScI), under contract with the National Aeronautics and Space Administration.
The MeerKAT data were processed using the MALS computing facility at IUCAA (https://mals.iucaa.in/releases).

\section*{Data Availability}

The data used in this work are obtained using MeerKAT (SSV-20180516-NG-02) and GMRT (ddtC003). The raw data will become available for public use in accordance with the observatory policy.  The MeerKAT data products will be publicly released through the survey website: https://mals.iucaa.in.

\def\aj{AJ}%
\def\actaa{Acta Astron.}%
\def\araa{ARA\&A}%
\def\apj{ApJ}%
\def\apjl{ApJ}%
\def\apjs{ApJS}%
\def\ao{Appl.~Opt.}%
\def\apss{Ap\&SS}%
\def\aap{A\&A}%
\def\aapr{A\&A~Rev.}%
\def\aaps{A\&AS}%
\def\azh{AZh}%
\def\baas{BAAS}%
\def\bac{Bull. astr. Inst. Czechosl.}%
\def\caa{Chinese Astron. Astrophys.}%
\def\cjaa{Chinese J. Astron. Astrophys.}%
\def\icarus{Icarus}%
\def\jcap{J. Cosmology Astropart. Phys.}%
\def\jrasc{JRASC}%
\def\mnras{MNRAS}%
\def\memras{MmRAS}%
\def\na{New A}%
\def\nar{New A Rev.}%
\def\pasa{PASA}%
\def\pra{Phys.~Rev.~A}%
\def\prb{Phys.~Rev.~B}%
\def\prc{Phys.~Rev.~C}%
\def\prd{Phys.~Rev.~D}%
\def\pre{Phys.~Rev.~E}%
\def\prl{Phys.~Rev.~Lett.}%
\def\pasp{PASP}%
\def\pasj{PASJ}%
\def\qjras{QJRAS}%
\def\rmxaa{Rev. Mexicana Astron. Astrofis.}%
\def\skytel{S\&T}%
\def\solphys{Sol.~Phys.}%
\def\sovast{Soviet~Ast.}%
\def\ssr{Space~Sci.~Rev.}%
\def\zap{ZAp}%
\def\nat{Nature}%
\def\iaucirc{IAU~Circ.}%
\def\aplett{Astrophys.~Lett.}%
\def\apspr{Astrophys.~Space~Phys.~Res.}%
\def\bain{Bull.~Astron.~Inst.~Netherlands}%
\def\fcp{Fund.~Cosmic~Phys.}%
\def\gca{Geochim.~Cosmochim.~Acta}%
\def\grl{Geophys.~Res.~Lett.}%
\def\jcp{J.~Chem.~Phys.}%
\def\jgr{J.~Geophys.~Res.}%
\def\jqsrt{J.~Quant.~Spec.~Radiat.~Transf.}%
\def\memsai{Mem.~Soc.~Astron.~Italiana}%
\def\nphysa{Nucl.~Phys.~A}%
\def\physrep{Phys.~Rep.}%
\def\physscr{Phys.~Scr}%
\def\planss{Planet.~Space~Sci.}%
\def\procspie{Proc.~SPIE}%
\let\astap=\aap
\let\apjlett=\apjl
\let\apjsupp=\apjs
\let\applopt=\ao

\bibliographystyle{mnras}
\bibliography{J2023}


\bsp	
\label{lastpage}
\end{document}